# Incorporation and effects of mesoporous SiO$_2$-CaO nanospheres loaded with ipriflavone on osteoblast/osteoclast cocultures

Laura Casarrubios[a], Natividad Gómez-Cerezo[b,c], María José Feito[a],

María Vallet-Regí[b,c,*], Daniel Arcos[b,c,*], María Teresa Portolés[a,*]

[a] *Departamento de Bioquímica y Biología Molecular, Facultad de Ciencias Químicas, Universidad Complutense de Madrid, Instituto de Investigación Sanitaria del Hospital Clínico San Carlos (IdISSC), 28040-Madrid, Spain.*

[b] *Departamento de Química en Ciencias Farmacéuticas, Facultad de Farmacia, Universidad Complutense de Madrid, Instituto de Investigación Sanitaria Hospital 12 de Octubre i+12, Plaza Ramón y Cajal s/n, 28040 Madrid, Spain.*

[c] *CIBER de Bioingeniería, Biomateriales y Nanomedicina, CIBER-BBN, Madrid, Spain.*

\* Corresponding authors

*E-mail address:* portoles@quim.ucm.es, arcosd@ucm.es, vallet@ucm.es

Phone: +34 91 394 4666; Fax: +34 91 394 4159

# Abstract

Mesoporous nanospheres in the system $SiO_2$-CaO (NanoMBGs) with a hollow core surrounded by a radial arrangement of mesopores were characterized, labeled with FITC (FITC-NanoMBGs) and loaded with ipriflavone (NanoMBG-IPs) in order to evaluate their incorporation and their effects on both osteoblasts and osteoclasts simultaneously and maintaining the communication with each other in coculture. The influence of these nanospheres on macrophage polarization towards pro-inflammatory M1 or reparative M2 phenotypes was also evaluated in basal and stimulated conditions through the expression of CD80 (as M1 marker) and CD206 (as M2 marker) by flow cytometry and confocal microscopy. NanoMBGs did not induce the macrophage polarization towards the M1 pro-inflammatory phenotype, favoring the M2 reparative phenotype and increasing the macrophage response capability against stimuli as LPS and IL-4. NanoMBG-IPs induced a significant decrease of osteoclast proliferation and resorption activity after 7 days in coculture with osteoblasts, without affecting osteoblast proliferation and viability. Drug release test demonstrated that only a fraction of the payload is released by diffusion, whereas the rest of the drug remains within the hollow core after 7 days, thus ensuring the local long-term pharmacological treatment beyond the initial fast IP release. All these data ensure an appropriate immune response to these nanospheres and the potential application of NanoMBG-IPs as local drug delivery system in osteoporotic patients.

**Keywords:** mesoporous bioactive glasses; nanospheres; osteoblasts; osteoclasts; ipriflavone.

# 1. Introduction

Throughout life, bone tissue undergoes continuous remodeling that requires the concerted action of bone-forming osteoblasts and bone-resorbing osteoclasts [1]. Osteoblasts are mononuclear cells that differentiate from bone-specific mesenchymal stem cells and carry out three important roles: the synthesis and secretion of most proteins of the bone extracellular matrix (ECM), the induction of ECM mineralization and the regulation of osteoclast differentiation for bone resorption [2-4]. On the other hand, osteoclasts are multinucleated giant cells that differentiate from hematopoietic stem cells and perform the bone resorption by the attachment to the bone surface and the secretion of hydrogen ions and lysosomal enzymes that degrade the bone ECM producing irregular cavities [5-7]. Osteoblasts and osteoclasts can communicate with each other through mechanisms that involve direct cell-cell contact, ECM interactions and the release of different cytokines [8]. In this context, osteoblasts express *M-Csf* and *Rankl* genes, which are the two main genes required for osteoclast differentiation. Therefore, it is possible to obtain osteoclast-like cells *in vitro* by culturing osteoclast progenitor cells in the presence of the proteins encoded by these genes: the macrophage/monocyte-colony forming factor (M-CSF) and the receptor activator of nuclear factor kappa-B ligand (RANKL), respectively [9,10]. Osteoblasts also produce osteoprotegerin (OPG), a soluble receptor that acts as a decoy receptor for RANKL and negatively regulates osteoclast differentiation [11]. The continuous stimulation of mononuclear pre-osteoclasts with M-CSF and RANKL induces the osteoclastogenesis by promoting the fusion of these cells into multinucleated osteoclasts and the formation of "ruffled membrane", critical for bone resorption [12,13]. During resorption process, the attachment of osteoclasts to the bone surface involves the creation of the "sealing zone", rich in F-actin, which forms a ring to isolate the resorptive space (resorption lacuna) from the surrounding bone [14,15].

Matrix-degrading enzymes (cathepsin K), hydrogen ions and chloride ions are released into the resorption lacunae by the ruffled membrane, formed by fusion of secretory vesicles into the plasma membrane within the sealing zone, producing acidification [16-18], the dissolution of the bone mineral component and the enzymatic degradation of the bone organic matrix [19,20]. The equilibrium between bone resorption by osteoclasts and bone formation by osteoblasts is necessary to maintain the structural skeleton integrity and mineral homeostasis [21]. The alterations of bone remodeling, that is influenced by mechanical, genetic, vascular, nutritional, hormonal and local factors, are involved in the pathogenesis of different skeletal diseases, including osteoporosis [22]. Antiresorptive and anabolic therapies, with different drugs and biomaterials, have been designed for the treatment of osteoporosis with the aim of activating bone formation or/and inhibiting osteoclast function and survival [23,24]. The *in vitro* evaluation of these potential treatments before their *in vivo* application, requires the design of experimental models, closer to physiological conditions, which allow osteoclasts and osteoblasts communicate with each other in the presence of drugs or/and biomaterials.

In the present study, monocultures and cocultures of osteoblasts and osteoclast-like cells were carried out to investigate *in vitro* the incorporation and the effects of mesoporous $SiO_2$-CaO nanospheres (NanoMBGs) loaded with ipriflavone (IP). NanoMBG nanospheres are intended for osteoporosis treatment by intraosseous administration, so that they would be in direct contact with bone cells. Besides, binary $SiO_2$-CaO compositions have widely demonstrated their bioactive behavior when tested with biological significant solutions [25]. IP is a synthetic isoflavone that inhibits bone resorption, maintains bone density and prevents osteoporosis [26]. Oral administration of 1,200 mg/day of IP in patients with primary hyperparathyroidism showed that ipriflavone is also indicated in the treatment of metabolic bone diseases characterized by a high bone turnover [27]. IP is metabolized by first pass metabolism, which leads to the low bioavailability and variation in blood concentration. For this reason, the administration of IP loaded into nanoparticles for intracellular release is a very interesting alternative, since the amount of IP required to affect the bone cells function would be much lower.

## 2. Materials and methods

*2.1. Preparation and characterization of mesoporous $SiO_2$-CaO nanospheres (NanoMBGs) and fluorescein isothiocyanate labeled nanospheres (FITC-NanoMBGs)*

The synthesis of hollow mesoporous $SiO_2$-CaO nanospheres (NanoMBGs) was carried out by the method described by Li et al. [28]. 80 mg of poly(styrene)-block-poly(acrylic acid) PS-b-PAA were dissolved in 16 ml of tetrahydrofurane (THF) at room temperature. This solution was subsequently poured in a hexadecyl trimethyl ammonium bromide (CTAB) solution, previously prepared by dissolving 160 mg of CTAB in 74 ml of deionized water and 2.4 ml of ammonia (28% w/v) and gently stirred in an incubator at 37ºC. The mixture was magnetically stirred for 20 minutes and a solution of 25 ml of TEP in 1.6 ml of ethanol was added drop by drop and stirred for another 20 minutes. Thereafter, a solution of 125 mg of $Ca(NO_3)·4H_2O$ in 1.6 ml of water was also added and stirred for 10 minutes and, finally the silica source was incorporated as a solution of 0.52 ml of TEOS in 1.6 ml of ethanol. After stirring for 24 hours, the product was collected by centrifugation at 10,000 rpm (g = 16.466) for 10 minutes and washed three times with a mixture of ethanol-water (50:50). The product was dried at 30 ºC under vacuum conditions and the organic template was removed by calcination at 550 ºC for 4 hours with a heating rate of 1 ºC $min^{-1}$. All reactants were purchased from Sigma- Aldrich (St. Louis, MO, USA).

For fluorescein isothiocyanate (FITC) labeling of NanoMBGs, 50 mg of nanospheres were degasified at 80ºC for 24 hours and resuspended in 4 ml of toluene. Besides, 44.3 µl of aminopropyl triethoxysiliane (APTES) were dissolved in 0.5 ml of ethanol and reacted with 0.6 mg of fluorescein isothiocyanate for 5 hours. This solution was added dropwise on the NanoMBG suspension and reacted at 80ºC for 12 hours under nitrogen atmosphere. Finally, fluorescein labelled NanoMBGs (FITC-NanoMBGs) were thoroughly washed and centrifuged several times at 10,000 rpm (g = 16.466) for 10 minutes to remove the excess of fluorescein non-covalently adsorbed to the nanospheres.

NanoMBGs were characterized by scanning electron microscopy (SEM) using a JEOL F-6335 microscope (JEOL Ltd., Tokyo, Japan), operating at 20 kV and equipped with an energy dispersive X-ray spectrometer (EDS). Previously, the samples were mounted on stubs and gold coated in vacuum using a sputter coater (Balzers SCD 004, Wiesbaden- Nordenstadt, Germany).

Transmission electron microscopy (TEM) was carried out using a JEOL-1400 microscope, operating at 300 kV (Cs 0.6mm, resolution 1.7 Å). Images were recorded using a CCD camera (model Keen view, SIS analyses size 1024 X 1024, pixel size 23.5mm x 23.5mm) at 60000X magnification using a low-dose condition.

Nitrogen adsorption/desorption isotherms were obtained with an ASAP 2020 porosimeter. NanoMBGs were previously degassed under vacuum for 15 h, at 150 ºC. The surface area was determined using the Brunauer-Emmett-Teller (BET) method. The pore size distribution between 0.5 and 40 nm was determined from the adsorption branch of the isotherm by means of the Barret-Joyner-Halenda (BJH) method. The surface area was calculated by the BET method and the pore size distribution was determined by the BJH method using the adsorption branch of the isotherm.

Fourier-transform infrared spectroscopy was done using a Nicolet Magma IR 550 spectrometer and using the attenuated total reflectance (ATR) sampling technique with a Golden Gate accessory.

Thermogravimetric analysis (TGA) was carried out using a TG/DTA Seiko SSC/5200 thermobalance (Seiko Instruments, Chiva, Japan) between 50 ºC and 600 ºC at a heating rate of 1 ºC min$^{-1}$, using aluminium crucibles and $\alpha$-Al$_2$O$_3$ as reference.

*2.2. Ipriflavone loading into NanoMBs (NanoMBG-IPs)*

Drug loading was carried out by dissolving 300 mg of IP (7-Isopropoxy-3-phenyl-4H-1-benzopyran-4-one, Sigma-Aldrich, St. Louis, MO, USA) in 6 ml of acetone. Thereafter 80 mg of NanoMBGs were suspended in this mixture and stirred for 24 hours to allow IP incorporation. Thereafter the NanoMBG-IP nanospheres were filtered under vacuum using a polyamide filter and thoroughly washed with water to remove the IP physiosorbed on nanosphere surface.

*2.3. Ipriflavone release from NanoMBG-IPs*

IP release test were carried out by placing 4 mg of NanoMBG-IPs in transwell inserts (0.4 μm pore size, Corning, USA) in 24 well culture plates. Due to the almost complete water insolubility of ipriflavone, a mixture of 2-propanol:water (60:40 v/v), already used in previous ipriflavone delivery assays [29-31] was chosen as release medium. Therefore, 1.5 ml of this mixture were added to the wells of the same plates in order to evaluate the drug release. UV–VIS spectrophotometry was employed for measurements of the drug concentration (Unicam UV – 500 UV-Visible1100 spectrophotometer) at 299 nm in quintuplicate.

*2.4. Culture of human Saos-2 osteoblasts*

Human Saos-2 osteoblasts (American Type Culture Collection, ATCC) were seeded in 6 well culture plates (Corning, USA), at a density of $10^5$ cells/ml, in 2 ml of Dulbecco's Modified Eagle Medium (DMEM) supplemented with 10% fetal bovine serum (FBS, Gibco, BRL), 1 mM L-glutamine (BioWhittaker Europe, Belgium), penicillin (200 μg/ml, BioWhittaker Europe, Belgium), and streptomycin (200 μg/ml, BioWhittaker Europe, Belgium) at 37 ºC under a $CO_2$ (5%) atmosphere. After 24 hours of culture in the presence or the absence of 50 μg/ml of NanoMBGs, the osteoblasts were washed with phosphate buffered saline (PBS), harvested using 0.25% trypsin-EDTA solution and counted with a Neubauer hemocytometer for the analysis of cell proliferation. Then, cells were centrifuged at 310 g for 10 min and resuspended in PBS for the analysis of cell cycle and apoptosis by flow cytometry as described below.

*2.5. Cell-cycle analysis and apoptosis detection by flow cytometry*

Cells were resuspended in PBS (0.5 ml) and incubated with 4.5 ml of ethanol 70% during 4 hours at 4 ºC. Then, cells were centrifuged at 310 g for 10 min, washed with PBS and resuspended in 0.5 ml of PBS with 0.1 % Triton X-100, 20 μg/ml of propidium iodide (IP) and 0.2 mg/ml of RNAsa (Sigma-Aldrich, St. Louis, MO, USA). After incubation at 37 ºC for 30 min, the fluorescence of PI was excited by a 15 mW laser tuning to 488 nm and the emitted fluorescence was measured with a 585/42 band pass filter in a FACScan Becton Dickinson flow cytometer. The cell percentage in each cycle phase: $G_0/G_1$, S and $G_2/M$ was calculated with the CellQuest Program of Becton Dickinson and the $SubG_1$ fraction was used as indicative of apoptosis. The conditions for data acquisition and analysis were established using negative and positive controls with the CellQuest Program of Becton Dickinson. These conditions were maintained during all the experiments. At least 10,000 cells were analyzed in each sample.

*2.6. Culture of RAW-264.7 macrophages*

RAW-264.7 cells (American Type Culture Collection, ATCC) were seeded in 6 well culture plates (Corning, USA), at a density of $10^5$ cells/ml, in 2 ml of Dulbecco's Modified Eagle Medium (DMEM) supplemented with 10% fetal bovine serum (FBS, Gibco, BRL), 1 mM L-glutamine (BioWhittaker Europe, Belgium), penicillin (200 µg/ml, BioWhittaker Europe, Belgium), and streptomycin (200 µg/ml, BioWhittaker Europe, Belgium) at 37 ºC under a $CO_2$ (5%) atmosphere. After 24 hours of culture in the presence or the absence of 50 µg/ml of NanoMBGs, the attached RAW-264.7 cells were washed with phosphate buffered saline (PBS), harvested using cell scrapers and counted with a Neubauer hemocytometer for the analysis of cell proliferation. Then, cells were centrifuged at 310 g for 10 min, resuspended in fresh medium for the analysis of viability, cell size and complexity by flow cytometry as described below.

*2.7. Cell viability, size and complexity analysis by flow cytometry*

Cell viability was evaluated by exclusion of propidium iodide (PI; 0.005 % wt/vol in PBS, Sigma-Aldrich, St. Louis, MO, USA). PI was added to the cell suspensions in order to stain the DNA of dead cells. The fluorescence of PI was excited by a 15 mW laser tuning to 488 nm and the emitted fluorescence was measured with a 530/30 band pass filter in a FACScalibur Becton Dickinson flow cytometer.

Forward angle (FSC) and side angle (SSC) scatters were evaluated as indicative of cell size and complexity, respectively, using a FACScalibur Becton Dickinson flow cytometer. The conditions for the data acquisition and analysis were established with the CellQuest Program of Becton Dickinson.

*2.8. Polarization of RAW-264.7 macrophages towards pro-inflammatory M1 and reparative M2 phenotypes. Flow cytometry and confocal microscopy studies*

To study the effect of NanoMBGs on macrophage polarization towards pro-inflammatory M1 and reparative M2 phenotypes, RAW-264.7 macrophages were cultured with 50 μg/ml of these nanospheres for 24 h in the presence of either *E. coli* lipopolysaccharide (LPS, 250 ng/ml, Sigma-Aldrich Corporation, St. Louis, MO, USA) or interleukin 4 (IL-4, 20 ng/ml, Sigma-Aldrich Corporation, St. Louis, MO, USA) as pro-inflammatory and reparative stimuli, respectively [32]. Controls without nanospheres and in the absence of stimuli were carried out in parallel. The expression of either CD80 as M1 marker [33] or CD206 as M2 marker [32] was quantified by flow cytometry after treatment with specific antibodies. With this objective, after detachment and centrifugation, cells were incubated in 45 μl of staining buffer (PBS, 2.5% FBS Gibco, BRL and 0.1% sodium azide, Sigma-Aldrich Corporation, St. Louis, MO, USA) with 5 μl of normal mouse serum inactivated for 15 min at 4 ºC in order to block the Fc receptors on the macrophage plasma membrane, before adding the primary antibody, and to prevent non-specific binding. Then, to quantify the CD80 and CD206 expression by flow cytometry, cells were incubated with either phycoerythrin (PE) conjugated anti-mouse CD80 antibody (2.5 μg/ml, BioLegend, San Diego, California) or fluorescein isothiocyanate (FITC) conjugated anti-mouse CD206 (2.5 μg/ml, BioLegend, San Diego, California) for 30 min in the dark. Labelled cells were analyzed using a FACSCalibur flow cytometer. PE fluorescence was excited at 488 nm and measured at 585/42 nm. FITC fluorescence was excited at 488 nm and measured with a 530/30 band pass filter. For confocal microscopy studies, macrophages cultured on glass coverslips were fixed with 3.7% paraformaldehyde (Sigma-Aldrich Corporation, St. Louis, MO, USA) in PBS for 10 min, washed with PBS and permeabilized with 0.1% Triton X-100 (Sigma-Aldrich Corporation, St. Louis, MO, USA) for 3 min. The samples were then washed with PBS and preincubated with PBS containing 1% BSA (Sigma-Aldrich Corporation, St. Louis, MO, USA) for 30 min to prevent non-specific binding. Samples were incubated in 1 ml of staining buffer with either phycoerythrin (PE) conjugated anti-mouse CD80 antibody (2.5 μg/ml, BioLegend, San Diego, California) or fluorescein isothiocyanate (FITC) conjugated anti-mouse CD206 (2.5 μg/ml, BioLegend, San Diego, California) for 30 min at 4ºC in the dark. Samples were then washed with PBS and the cell nuclei were stained with 3 μM DAPI (4′-6-diamidino-2′-phenylindole, Molecular Probes) for 5 min. Samples were examined using a Leica SP2 Confocal Laser Scanning Microscope. PE fluorescence was excited at 488 nm and measured at 575-675 nm. FITC fluorescence was excited at 488 nm and measured at 491-586 nm. DAPI fluorescence was excited at 405 nm and measured at 420–480 nm.

*2.9. Osteoclast differentiation from murine RAW 264.7 macrophages. Confocal microscopy studies*

Murine RAW-264.7 macrophages (2 x $10^4$ cells/ml) were seeded on glass coverslips and cultured in the presence of 50 μg/ml of NanoMBGs in Dulbecco's Modified Eagle Medium (DMEM), supplemented with 10% fetal bovine serum (FBS, Gibco, BRL), 1 mM L-glutamine (BioWhittaker Europe, Belgium), penicillin (200 μg/ml, BioWhittaker Europe, Belgium), and streptomycin (200 μg/ml, BioWhittaker Europe, Belgium). To stimulate osteoclast differentiation, 40 ng/ml of mouse RANK Ligand recombinant protein (TRANCE/RANKL, carrier-free, BioLegend, San Diego) and 25 ng/ml recombinant human macrophage-colony stimulating factor (M-CSF, Milipore, Temecula) were added to the culture medium. It has been demonstrated that M-CSF modulates multiple steps of human osteoclastogenesis and osteoclast-resorbing activity, but is not required for osteoclast survival [34]. Osteoclast precursors recognize RANKL through cell-to-cell interactions with osteoblasts and differentiate into osteoclasts in the presence of M-CSF. RANKL also stimulates the survival and bone-resorbing activity of osteoclasts [35]. In our experimental conditions, we have previously shown that M-CSF and RANKL induce the differentiation from RAW-264.7 macrophages into osteoclasts by stimulating their fusion as precursor cells and their osteoclast-resorbing activity [9,10]. Cells were cultured under a 5% $CO_2$ atmosphere and at 37ºC for 7 days. Controls in the absence of nanospheres were carried out in parallel. For confocal microscopy studies, cells were washed with PBS, fixed with 3.7% paraformaldehyde in PBS for 10 min, permeabilizated with 0.1% Triton X-100 for 3 min and preincubated with PBS containing 1% BSA for 30 min. Then, cells were incubated with rhodamine phalloidin (1:40, v/v Molecular Probes) for 20 min to stain F-actin filaments. Samples were then washed with PBS and cell nuclei were stained with 3 μM DAPI (4′-6-diamidino-2′-phenylindole; Molecular Probes) for 5 min. After mounting with Prolong Gold reagent (Thermo Fisher Scientific), cells were examined using a Leica SP2 Confocal Laser Scanning Microscope. Rhodamine fluorescence was excited at 540 nm and measured at 565 nm. DAPI fluorescence was excited at 405 nm and measured at 420–480 nm.

*2.10. Osteoclast resorption activity. Scanning electron microscopy studies*

To evaluate the resorption activity of osteoclasts, RAW-264.7 macrophages were seeded on the surface of nanocrystalline hydroxyapatite (nano-HA) disks and differentiate into osteoclasts in the presence of 50 μg/ml of NanoMBGs as it is described above. Nano-HA disks were prepared by controlled precipitation of calcium and phosphate salts and subsequently heated at temperatures below the sintering point, as previously described by our research group [9]. Controls in the absence of nanospheres were carried out in parallel. After 7 days of differentiation, cells were detached using cell scrapers and disks were dehydrated, coated with gold-palladium and examined with a JEOL JSM-6400 scanning electron microscope in order to observe the geometry of resorption cavities produced by osteoclasts on the surface of nano-HA disks.

*2.11. Osteoblast/osteoclast coculture*

RAW-264.7 cells (American Type Culture Collection, ATCC) were seeded in 6 well culture plates (Corning, USA), at a density of $2 \times 10^4$ cells/ml, in 2.3 ml of Dulbecco's Modified Eagle Medium (DMEM) supplemented with 10% fetal bovine serum (FBS, Gibco, BRL), 1 mM L-glutamine (BioWhittaker Europe, Belgium), penicillin (200 μg/ml, BioWhittaker Europe, Belgium), and streptomycin (200 μg/ml, BioWhittaker Europe, Belgium). In order to stimulate osteoclast-like cell differentiation, 40 ng/ml of mouse RANK Ligand recombinant protein (TRANCE/RANKL, carrier-free, BioLegend, San Diego) and 25 ng/ml recombinant human macrophage-colony stimulating factor (M-CSF, Millipore, Temecula) were added to the culture medium. Simultaneously, Saos-2 osteoblasts were seeded at a density of $2 \times 10^4$ cells/ml in transwell inserts (0.4 μm pore size, Corning, USA) in 1.3 ml of the same culture medium and placed into the 6 well culture plates containing seeded RAW-264.7 cells (Scheme 1). This cocultures were carried out in the presence or the absence of 50 μg/ml of NanoMBGs without or with ipriflavone (NanoMBG-IPs) for 7 days at 37 ºC under a $CO_2$ (5%) atmosphere. Separately, RAW-264.7 macrophages and Saos-2 osteoblasts were cultured alone in wells and transwell inserts, respectively as controls, to compare the results obtained with NanoMBGs and NanoMBG-IPs on these two cell types in coculture and separately without cell communication. For this reason we maintained RANKL at the same dose for monoculture and coculture. After co-culturing, osteoblasts and osteoclast were washed with phosphate buffered saline (PBS), harvested using 0.25% trypsin-EDTA solution and cell scrapers respectively, and counted with a Neubauer hemocytometer for the analysis of cell number. Then, cells were centrifuged at 310 g for 10 min and resuspended in PBS for the analysis of cell cycle and viability by flow cytometry as described above. For confocal microscopy studies of cocultured osteoblasts and osteoclasts, Saos-2 osteoblasts and RAW-264.7 cells were seeded on glass coverslips into transwell inserts and wells respectively, and cocultured as described above. After 7 days of coculture, cells were washed with PBS, fixed with 3.7% paraformaldehyde in PBS for 10 min, permeabilized with 0.1% Triton X-100 for 3 min and preincubated with PBS containing 1% BSA for 30 min. Then, cells were incubated with rhodamine phalloidin (1:40, v/v Molecular Probes) for 20 min to stain F-actin filaments. Samples were then washed with PBS and cell nuclei were stained with 3 µM DAPI (4′-6-diamidino-2′-phenylindole; Molecular Probes) for 5 min. After mounting with Prolong Gold reagent (Thermo Fisher Scientific), cells were examined using a Leica SP2 Confocal Laser Scanning Microscope. Rhodamine fluorescence was excited at 540 nm and measured at 565 nm. DAPI fluorescence was excited at 405 nm and measured at 420–480 nm.

To evaluate the resorption activity of osteoclasts in coculture, RAW-264.7 macrophages were seeded on the surface of nano-HA disks into 6 well culture plates with differentiation medium (with RANKL and M-CSF) and cocultured with Saos-2 osteoblasts previously seeded in the same well around nano-HA disks (Scheme 2).

*2.12. Incorporation of FITC-NanoMBGs by osteoblasts and osteoclasts*

The incorporation of FITC-NanoMBGs by osteoblasts and osteoclasts was quantified by flow cytometry after 3 days and 7 days of monoculture and coculture of these cells with 50 µg/ml of FITC-NanoMBGs under the conditions described above. Osteoblasts and osteoclast were washed with phosphate buffered saline (PBS), harvested using 0.25% trypsin-EDTA solution and cell scrapers respectively, centrifuged at 310 g for 10 min and resuspended in PBS for the analysis of FITC-NanoMBG incorporation by flow cytometry. FITC fluorescence was excited by a 15 mW laser tuning to 488 nm and measured with a 530/30 band pass filter in a FACScalibur Becton Dickinson Flow Cytometer. For confocal microscopy studies, cells were seeded on glass coverslips and cultured under the conditions described above. After 7 days of culture, cells were washed with PBS, fixed with 3.7% paraformaldehyde in PBS for 10 min, permeabilized with 0.1% Triton X-100 for 3 min and preincubated with PBS containing 1% BSA for 30 min. Then, cells were incubated with rhodamine phalloidin (1:40, v/v Molecular Probes) for 20 min to stain F-actin filaments. Samples were then washed with PBS and cell nuclei were stained with 3 µM DAPI (4′-6-diamidino-2′-phenylindole; Molecular Probes) for 5 min. After mounting with Prolong Gold reagent (Thermo Fisher Scientific), cells were examined using a Leica SP2 Confocal Laser Scanning Microscope. Rhodamine fluorescence was excited at 540 nm and measured at 565 nm. FITC fluorescence was excited at 488 nm and measured at 491-586 nm. DAPI fluorescence was excited at 405 nm and measured at 420–480 nm.

*2.13. Statistics*

Data are expressed as means + standard deviations of a representative of three repetitive experiments carried out in triplicate. Statistical analysis was performed by using the Statistical Package for the Social Sciences (SPSS) version 22 software. Statistical comparisons were made by analysis of variance (ANOVA). Scheffé test was used for *post hoc* evaluations of differences among groups. In all statistical evaluations, $p < 0.05$ was considered as statistically significant.

# 3. Results and discussion

*3.1. Characterization of NanoMBGs and ipriflavone release test*

NanoMBGs were morphologically characterized by SEM and TEM. SEM micrographs (Figure 1 a) show that the NanoMBG material consists on monodisperse spherical nanoparticles of around 250 nm in diameter, although some nanoparticles have certain degree of polyhedral morphology. This fact would be a consequence of the use of CTAB as secondary template, as this surfactant is placed at the external location of the particles and often leads to hexagonal polyhedral morphologies [36]. TEM images show that NanoMBGs exhibit a hollow core-shell structure due to the double-template method used in this work (Figure 1 b). Higher magnification (Figure 1 c) reveals that the shell is organized into a radial mesoporous structure from the hollow core towards the external surface.

Nitrogen adsorption/desorption isotherms (Figure 1 d) agree with this dual mesoporous structure. The adsorption curve corresponds to a type IV isotherm characteristic of mesoporous materials and the pore size is a monodispersed distribution centered at 2.3 nm, which would correspond to the mesopores of the outer shell. The size of the hollow core (about 200 nm) is too large to be measured by nitrogen adsorption and the pore size distribution could not be obtained. However, the H2 type hysteresis loop obtained with the desorption isotherm reveals a second kind of porosity with ink bottle morphology, which would indicate that the smaller pores of the shell are connected to the larger central pore of NanoMBGs.

Ipriflavone (IP) was loaded into NanoMBGs obtaining values of 13% in weight determined by thermogravimetric analysis. The chemical compositions of both NanoMBG and NanoMBG-IP nanospheres were determined by EDX spectroscopy during TEM observations and are also shown in Table 1. Despite of the incorporation of a phosphorous precursor as TEP during the synthesis, this element could not be detected by EDX, whereas silicon and calcium contents are very similar respect to the theoretical values. However, we decided to maintain the addition of TEP as we could observe that the presence of this reactive increased the content of Ca in NanoMBGs. In the absence of TEP, the calcium content in our nanoparticles could not overcome the 12 % value. Interestingly, NanoMBG-IPs keep the calcium content after the drug loading, which is a condition required for maintaining the bioactive behavior of mesoporous bioactive glasses [37,38]. This is possible due to the non-polar character of acetone used for the IP loading, which would avoid the calcium release commonly occurred when MBGs are soaked in aqueous mediums.

Finally, the textural parameters obtained by $N_2$ adsorption are shown in Table 1 and evidence that the high surface area and porosity NanoMBGs are dramatically decreased after IP incorporation.

Figure 2 a shows the FTIR spectra of NanoMBGs before and after loaded with IP (NanoMBG-IP), evidencing the incorporation of the drug with the appearance of the absorption bands corresponding to IP chemical groups. The drug delivery test (Figure 2 b) shows a fast release of the 18% of the IP during the first 10 hours, followed by a slower release of an additional 6 % of drug before reaching an asymptotic behavior. The test was followed for 7 days without observing further IP release to the medium. Since the release test was carried out in isopropanol:water media, these results cannot be extrapolated to physiological conditions, where IP is almost insoluble Therefore, it should be considered just as a proof of the capability of nanoMBG for loading and release IP in a medium of low polarity. It must be pointed that, even using a release media that facilitates IP release, most of IP is still retained within the core of NanoMBG nanospheres after 7 days, as could be confirmed by FTIR spectroscopy carried out on the particles collected after the drug release test (see Figure 2 a, NanoMBG-IP, 7 days). This behavior indicates that NanoMBG-IPs release a minor fraction of their payload by diffusion through the radial porosity of the shell, whereas most of the drug would be retained within the hollow core. In this sense, the total release of IP would be conditioned to nanoparticles degradation, thus ensuring the local long-term pharmacological treatment beyond the initial fast IP release. The degradation of MBGs has been widely studied and is initialized by an intense ionic exchange with the surrounding fluids [37,38]. In fact, this ionic exchange is the basis for their bone regenerative properties. It can be described as a sequence of reactions involving:

a) $Ca^{2+}$ by $H^+$ ionic exchange between the material and the surrounding media.

b) Release of soluble $SiO_2$ oligomers from the material to the surrounding fluid.

c) In the case of solutions with high calcium and phosphorous content (like the human plasma or cell culture media), a calcium phosphate phase very similar to the mineral component of the bone nucleates on the materials surface, thus ensuring the osteointegration of the MBGs with the hosting bone tissue.

*3.2. Effects of NanoMBGs on human Saos-2 osteoblasts*

Figure 3 shows the effects of NanoMBGs on proliferation (a), cell cycle profile (b) and cell cycle phases (c) of human Saos-2 osteoblasts cultured in the absence (white) or in the presence (grey) of 50 µg/ml of nanospheres for 24 hours. Saos-2 cells is an osteosarcoma cell line commonly used for *in vitro* evaluation of biomaterials designed for bone tissue due to its osteoblastic properties as production of mineralized matrix, high alkaline phosphatase levels, PTH receptors and osteonectin presence [39]. The analysis of the cell cycle by flow cytometry allows us to detect the percentage of cells in the progressive stages: $G_0/G_1$ phase (Quiescence/Gap1), S phase (Synthesis) and finally $G_2/M$ phase (Gap2 and Mitosis). This analysis also indicates the percentage of apoptotic cells with fragmented DNA corresponding to the $SubG_1$ fraction.

The treatment with these nanospheres produced a significant S phase increase ($p < 0.005$) and a significant $G_2/M$ phase decrease ($p < 0.05$) after 24 h of treatment (Figure 3 c) that could be related to the observed, but no significant decrease detected in osteoblast number (proliferation, Figure 3 a) and $G_0/G_1$ phase (Figure 3 c). On the other hand, although this nanomaterial induced a significant increase of $SubG_1$ fraction ($p<0.005$, Figure 3 c), very low levels of apoptosis were detected either in the absence or in the presence of these nanospheres.

*3.3. Effects of NanoMBGs on RAW-264.7 macrophages*

In order to know the effects of NanoMBG nanospheres on proliferation, viability, cell size and complexity of RAW-264.7 macrophages, these cells were cultured in the absence or in the presence of 50 μg/ml of nanospheres for 24 hours. Figure 4 shows that these nanoparticles did not induced significant alterations on cell proliferation, cell size and viability. However, a significant increase ($p < 0.005$) of cell complexity evaluated through the side angle scatter (SSC) was detected after the treatment with this nanomaterial. SSC is a special parameter of flow cytometry which can reflect the physical properties of a cell examined by the flow cytometer. It is a part of deflected laser light by the cell. The extent to which light scatters depends on the physical properties of a cell as its size and internal complexity. Factors that affect light scattering are the plasma membrane, cytoplasm, nucleus, mitochondria, pinocytic vesicles, lysosomes, and any granular material inside the cell. Cell shape and surface topography also contribute to the total light scattering [40,41]. Thus, the changes observed on SSC in Figure 4 could be due to the nanospheres uptake by macrophages.

The expression of either CD80 as M1 marker [33] or CD206 as M2 marker [32] after treatment with PE conjugated anti-mouse CD80 antibody or FITC conjugated anti-mouse CD206 antibody is shown in Figures 5 a, b, c, d, f and g and observed by confocal microscopy (Figures 5 e and h).

As it can be observed in Figures 5 a and b, the treatment with NanoMBGs in basal conditions (without stimulus) did not induce significant changes of M1 percentage and CD80 expression (M1 fluorescence intensity) but produced a significant increase of M2 percentage ($p<0.05$) accompanied by a significant decrease of CD206 fluorescence intensity (M2, $p<0.01$). This M2 fluorescence intensity corresponds to the mean fluorescence of the CD206$^+$ population (M2). This M2 population presented a slight but significant increase of cell number after treatment with nanospheres but, probably due to this fact, the mean fluorescence of this population shifts to a lower value. When macrophages were treated with LPS as inflammatory stimulus (Figures 5 c and d), these nanospheres did not modify M1 percentage but increased CD80 fluorescence intensity (M1, $p<0.005$), evidencing a higher response capability against the LPS stimulus in the presence of material. On the other hand, when macrophages were treated with IL-4 as reparative stimulus (Figures 5 f and g), NanoMBGs induced significant increases of both M2 percentage and CD206 expression (M2 fluorescence intensity, $p<0.005$). These results evidence that NanoMBGs did not induce the macrophage polarization towards the M1 pro-inflammatory phenotype in basal conditions, promoting the control of the M1/M2 balance with a slight shift towards M2 reparative phenotype and increasing the response capability against both stimuli LPS and IL-4. All these data ensure an appropriate immune response to these nanospheres. On the other hand, the macrophage phenotype polarization has been related to changes in cell shape [42] that were also observed in the present study when macrophages were stimulated with either LPS or IL-4 as pro-inflammatory and reparative stimulus, respectively. Thus, macrophages polarized towards the M1 (CD80$^+$) phenotype with LPS showed more spherical shape (Figure 5 e) than cells polarized towards the M2 (CD206$^+$) phenotype with IL-4 showing an elongated shape (Figure 5 h), in agreement with other authors [42] and previous studies [43].

*3.4. Incorporation FITC-NanoMBGs by osteoclasts and osteoblasts in monoculture and in coculture with each other.*

Figure 6 shows the incorporation of NanoMBGs labelled with FITC by osteoclasts and osteoblasts in monoculture (OC and OB) and in coculture (OC-co and OB-co) with each other. The percentage of cells with intracellular FITC-NanoMBGs and the fluorescence intensity of these cells were quantified by flow cytometry after 3 days (Figure 6 a) and 7 days (Figure 6 b). Confocal microscopy studies were carried out to observe these nanospheres incorporated by osteoclasts (Figure 6 d) and osteoblasts (Figure 6 f) in coculture after 7 days. Control cocultures of osteoclasts (Figure 6 c) with osteoblasts (Figure 6 e) in the absence of nanospheres were carried out in parallel.

As it can be observed in Figure 6 b, the percentage of cells with FITC-NanoMBGs and the fluorescence intensity of these cells were significantly higher in monocultured and cocultured osteoblasts than in monocultured and cocultured osteoclasts after 7 days of treatment ($p<0.005$). Confocal microscopy studies evidenced that after FITC-NanoMBG incorporation, osteoclasts and osteoblasts maintained their characteristic morphology in both monocultures (data not shown) and cocultures (Figures 6 d and f, respectively) in comparison with control monocultures (data not shown) and cocultures (Figures 6 c and e, respectively). These results evidenced that FITC-NanoMBG incorporation did not induced alterations in osteoclast and osteoblast morphology, allowing the osteoclastogenesis process that was confirmed by the observation of multinucleated cells with the F-actin ring (Figure 6 d), that allows creation of the "sealing zone", critical for bone resorption [14,15].

Figure 7 shows the cell number (7 a) and cell viability (7 b) of osteoclasts (OC) and osteoblasts (OB) in coculture with each other after 7 days of treatment with 50 μg/ml of either NanoMBG (OC-co and OB-co) or NanoMBG-IP (Oc-co/IP and OB-co/IP) nanospheres. As it can be observed, the presence of ipriflavone induced a significant decrease of osteoclast number ($p<0.005$) without affecting the viability of this cell type. On the other hand, ipriflavone did not induce changes on osteoblast proliferation and viability. The cell morphology of osteoclasts and osteoblasts in coculture in the presence of nanospheres without IP (7 c and 7 e, respectively) or with IP (7 d and 7 f, respectively) did not show alterations.

Figure 8 shows the resorption cavities left by osteoclasts on nanocrystalline hydroxyapatite disks after 7 days of differentiation in monoculture (a), in coculture with osteoblasts (b), in coculture with osteoblasts in the presence of 50 μg/ml of NanoMBGs (c and d) and in coculture with osteoblasts in the presence of 50 μg/ml of NanoMBG-IPs (e and f). As it can be observed, the dimensions of these cavities significantly decreased when osteoclasts are cocultured with osteoblasts (Figure 8 b) in comparison with the cavities left by osteoclasts in monoculture (Figure 8 a), evidencing that the presence of osteoblasts modulates the resorption activity of osteoclasts. These results could be due to the osteoblast production of osteoprotegerin (OPG), a soluble receptor that acts as a decoy receptor for RANKL and negatively regulates osteoclast differentiation [11]. The cavities left by osteoclasts in coculture and in the presence of NanoMBGs without ipriflavone (IP) (Figures 8 c and d) were similar to those obtained in the absence of spheres (Figure 8 b). However, in the presence of IP-loaded NanoMBGs, osteoclasts in coculture exhibited a lower resorptive activity and superficial erosion marks were observed (Figures 8 e and f), probably due to the IP action in agreement with the mechanisms of this agent [44]. Different authors have demonstrated that the bone-resorbing activity of osteoclasts is regulated by $Ca^{2+}$ levels and that high concentrations of this cation produce osteoclast retraction and dissipation of sealing zone, decreasing the bone resorption process [45,46].

Concerning IP mechanism, it has been described that IP inhibits the fusion of osteoclast precursor cells, bone resorption and tartrate-resistant acid phosphatase activity. These effects are mediated by specific IP receptors that induce a rapid increase in intracellular [$Ca^{2+}$] followed by a sustained elevation of this cation in osteoclasts and their precursor cells [44]. It must be highlighted that only a minor fraction of IP is released after 7 days from NanoMBG-IPs when soaked in isopropanol: water solution, in which IP is highly soluble. In cell culture media, the supposedly smaller amount of IP released seems to be enough to decrease the resorptive activity of osteoclasts. We envision that this behavior could be beneficial for the local delivery treatment of osteoporotic bone. Osteoporosis requires prolonged periods of drug administration. In the case of IP, the bioactive dosage strongly varies depending on the administration way. For instance, IP biodisponibility in plasma after oral administration is only 20% and around 120 mg per day of ipriflavone is required in plasma. The loading rate of IP in NanoMBGs is 13% and the *in vitro* release test suggests that only a minor fraction of the payload by diffusion through the radial pores of the shell. However, in contact with bone cells, this payload seems to be enough to initially decrease the resorptive activity of osteoclasts.

## 4. Conclusions

Mesoporous bioactive nanospheres in the system $SiO_2$-CaO (NanoMBGs) have been prepared via dual template method. NanoMBGs exhibit a hollow core and a radial mesoporous arrangement at the shell. The porous structure of these nanospheres allowed the loading of ipriflavone obtaining NanoMBG-IPs. However, ipriflavone is a highly insoluble compound in aqueous environment like culture media or human plasma. Even in low-polar conditions, where IP is highly soluble, most of the drug remains retained within the nanoparticles core. This fact would suggest that, under less favorable hydric physiological conditions, IP would be kept for longer periods, and it would enable the local long term pharmacological treatment.

NanoMBGs did not induce the macrophage polarization towards the M1 pro-inflammatory phenotype in basal conditions, promoting the control of the M1/M2 balance with a slight shift towards M2 reparative phenotype, increasing the response capability against stimuli (LPS and IL-4) and ensuring an appropriate immune response. The intracellular incorporation of FITC-NanoMBG after 7 days, detected in monocultures and cocultures of osteoblasts and osteoclasts, was significantly higher in osteoblasts. NanoMBG-IPs induced a significant decrease of osteoclast cell number and resorption activity in coculture with osteoblasts without affecting osteoblast proliferation and viability. All these data ensure an appropriate immune response to these nanospheres and the potential application of NanoMBG-IPs as local drug delivery system in osteoporotic patients. Further studies will be performed with *in vivo* models in order to determine the advantages of these NanoMBG-IPs to other anti-osteoporotic treatments and these results will be included in a future manuscript.

## Acknowledgements

This study was supported by research grants from Spanish MINECO (MAT2015-64831-R, MAT2016-75611-R AEI/FEDER, UE). MVR acknowledges funding from the European Research Council (Advanced Grant VERDI; ERC-2015-AdG Proposal No.694160). LC acknowledges the financial support from Comunidad de Madrid (Spain, CT4/17/CT5/17/PEJD-2016/BMD-2749). NGC is greatly indebted to Ministerio de Ciencia e Innovación for her predoctoral fellowship. The authors wish to thank the ICTS

**SCHEMES**

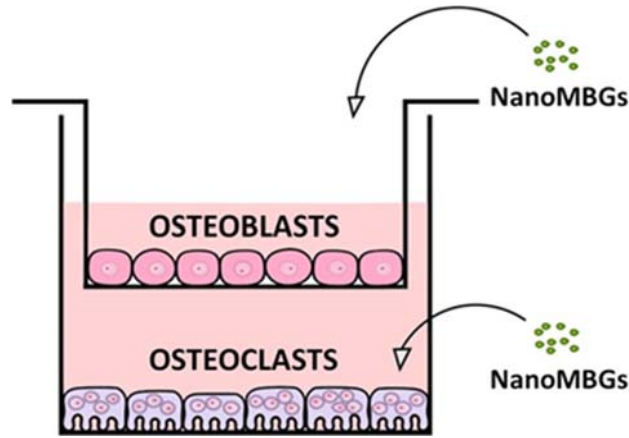

**Scheme 1.** Cocultures of human Saos-2 osteoblasts and osteoclast-like cells differentiated from RAW-264.7 macrophages in the presence of RANKL and M-CSF were carried to evaluate the incorporation and effects of 50 μg/ml of NanoMBG nanospheres with or without ipriflavone.

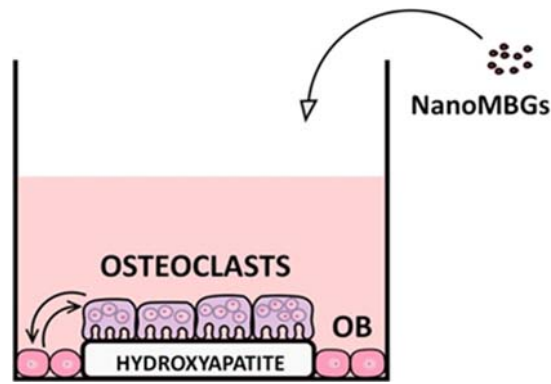

**Scheme 2.** Cocultures of human Saos-2 osteoblasts (OB) and osteoclast-like cells differentiated from RAW-264.7 macrophages in the presence of RANKL and M-CSF on nanocrystalline hydroxyapatite disks were carried to evaluate the effects of 50 μg/ml of NanoMBG nanospheres with or without ipriflavone on the resorption activity of osteoclasts.

# TABLES

**Table 1.** Chemical composition and textural properties of NanoMBG nanospheres before and after loading with ipriflavone. Values in brackets correspond to the theoretical chemical composition.

| Sample | Si (% atom) | Ca (% atom) | P (% atom) | Surface area (m$^2$·g$^{-1}$) | Porosity (cm$^3$g$^{-1}$) | Pore size (nm) |
|---|---|---|---|---|---|---|
| **NanoMBG** | 81.44 (79.4) | 18.56 (18.1) | - (2.5) | 543.6 | 0.435 | 2.2 |
| **NanoMBG-IP** | 83.36 | 16.63 | - | 14.2 | 0.057 | NA |

**FIGURES**

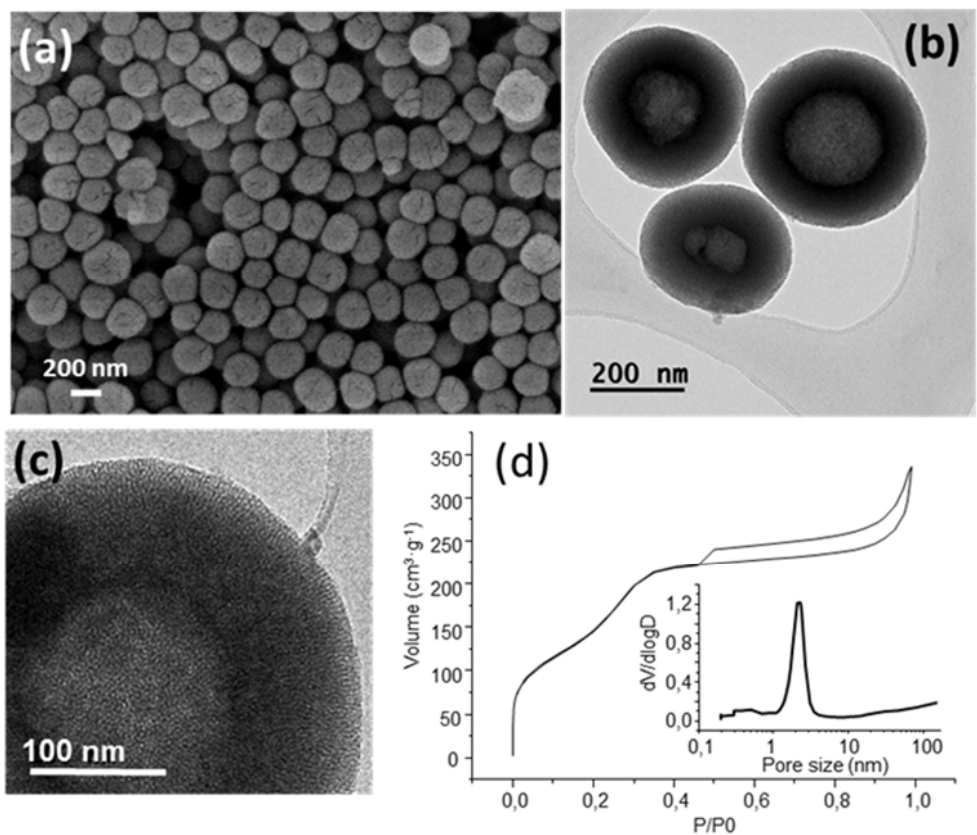

**Figure 1.** Scanning electron micrograph of NanoMBG nanospheres (**a**), high resolution transmission electron microscopy image of NanoMBG nanospheres (**b** and **c**), adsorption-desorption isotherms and pore size distribution (**d**).

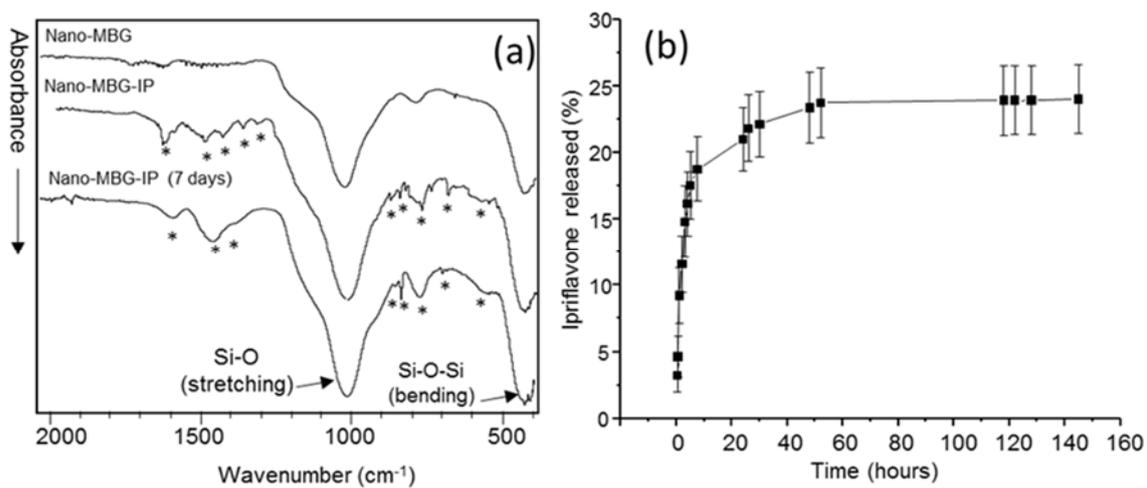

**Figure 2.** FTIR spectra of NanoMBG and NanoMBG-IP nanospheres before and after seven days of drug release test. * points absorption bands corresponding to the different functional groups of ipriflavone (**a**). Ipriflavone released as a function of soaking time (**b**).

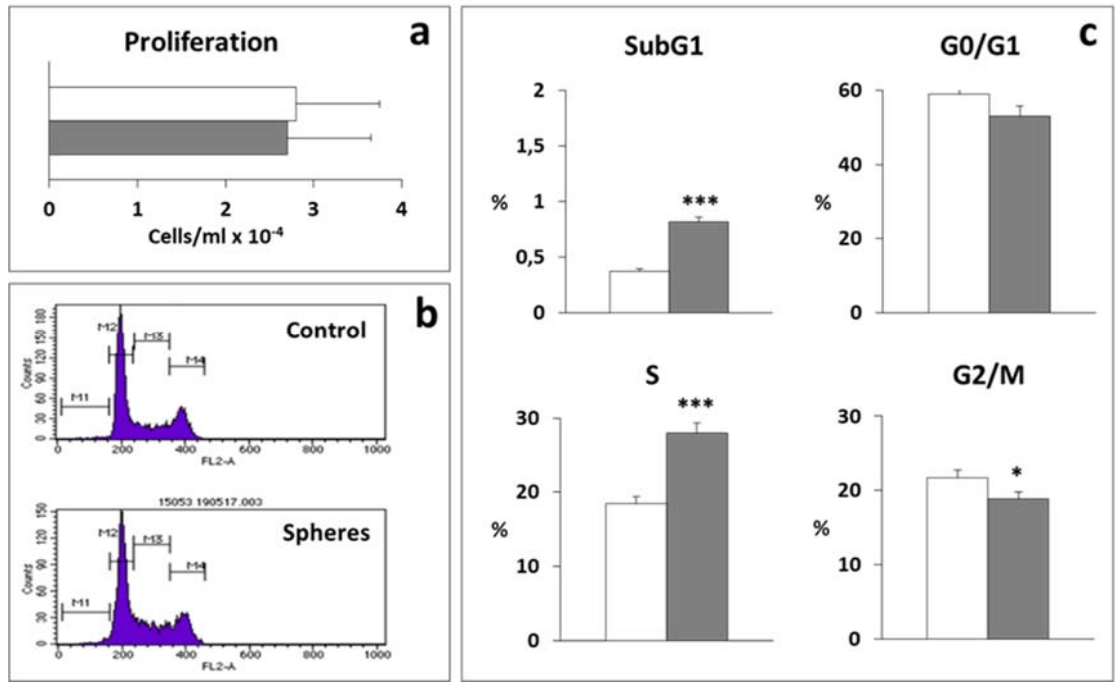

**Figure 3.** Effects of NanoMBG nanospheres on proliferation and cell cycle of human Saos-2 osteoblasts: **a)** proliferation, **b)** cell cycle profiles and **c)** cell cycle phases of cells cultured in the absence (white) or in the presence (grey) of 50 μg/ml of nanospheres for 24 hours. Statistical significance: *p < 0.05; *** p < 0.005.

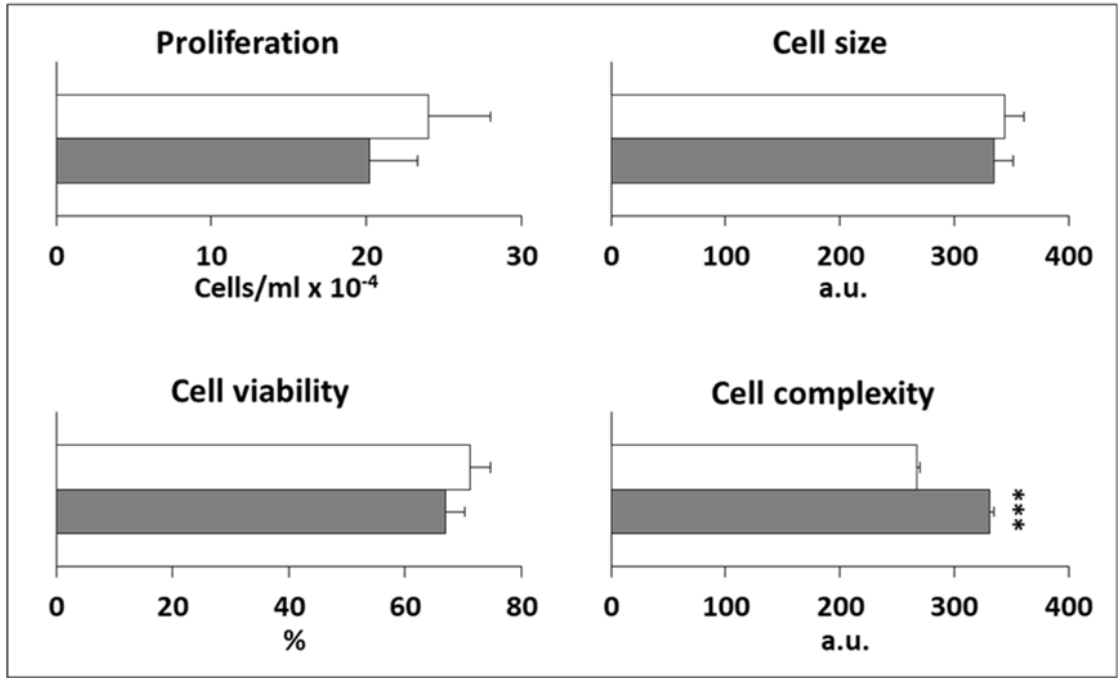

**Figure 4.** Effects of NanoMBG nanospheres on proliferation, cell viability, size and complexity of RAW-264.7 macrophages cultured in the absence (white) or in the presence (grey) of 50 μg/ml of nanospheres for 24 hours. Statistical significance: *** p < 0.005.

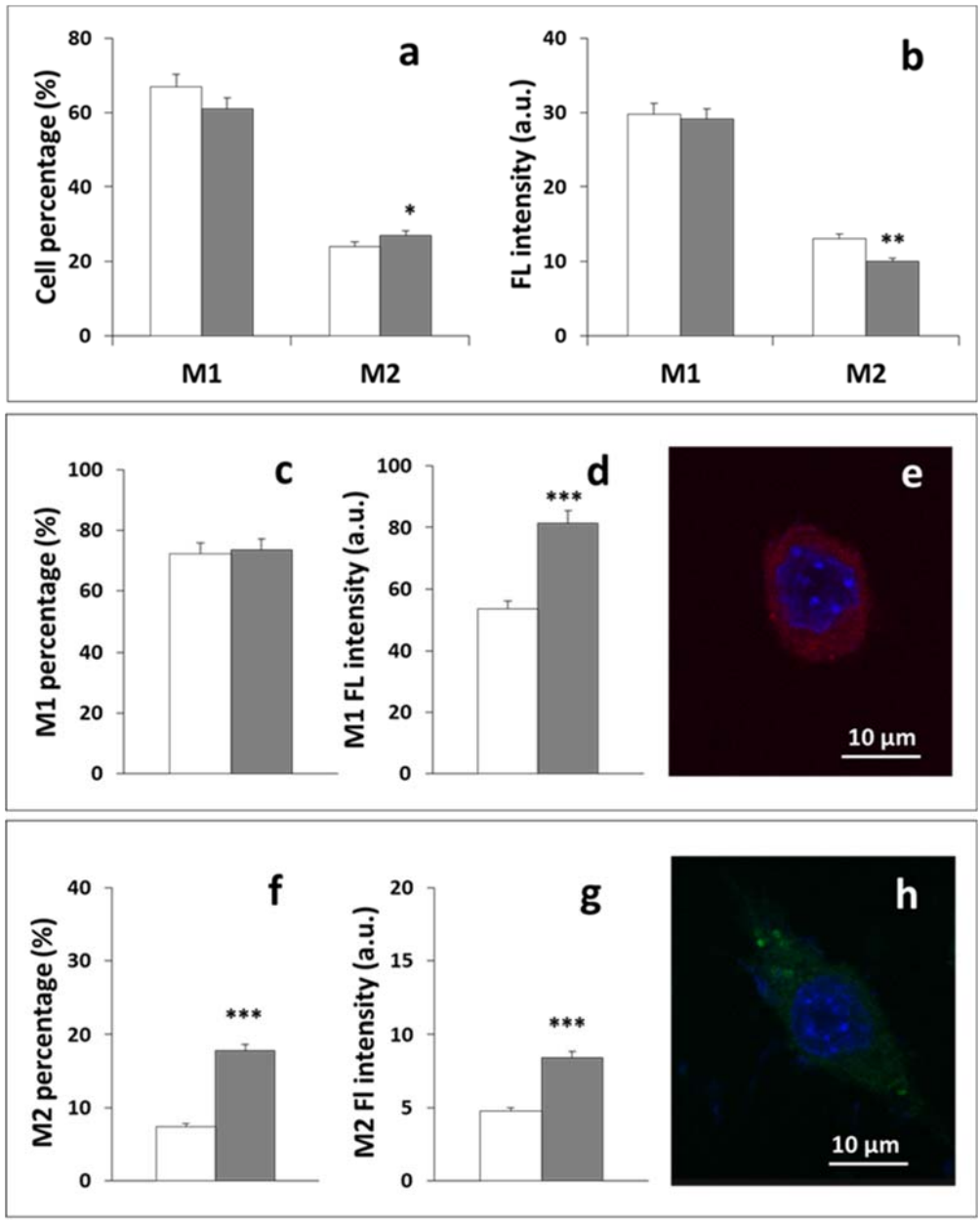

**Figure 5.** Effects of NanoMBG nanospheres on pro-inflammatory M1 and reparative M2 phenotypes of RAW-264.7 macrophages after 24 hours of treatment without stimulus (**a** and **b**) or with *E. coli* lipopolysaccharide (LPS) (**c**, **d** and **e**) as inflammatory stimulus or with interleukin 4 (IL-4) (**f**, **g** and **h**) as reparative stimulus. Cells were cultured in the absence (white) or in the presence (grey) of 50 μg/ml of nanospheres for 24 hours. Statistical significance: * $p < 0.05$, ** $p < 0.01$, *** $p < 0.005$. Confocal images show the CD80 (**e**) and CD206 (**h**) expression of M1 and M2 RAW-264.7 macrophages after treatment with LPS or IL-4 respectively, and in the presence of 50 μg/ml of nanospheres for 24 hours. CD80 (red) was detected with PE conjugated anti-mouse CD80 antibody, CD206 (green) was detected with FITC conjugated anti-mouse CD206 antibody, nuclei were stained with DAPI (blue).

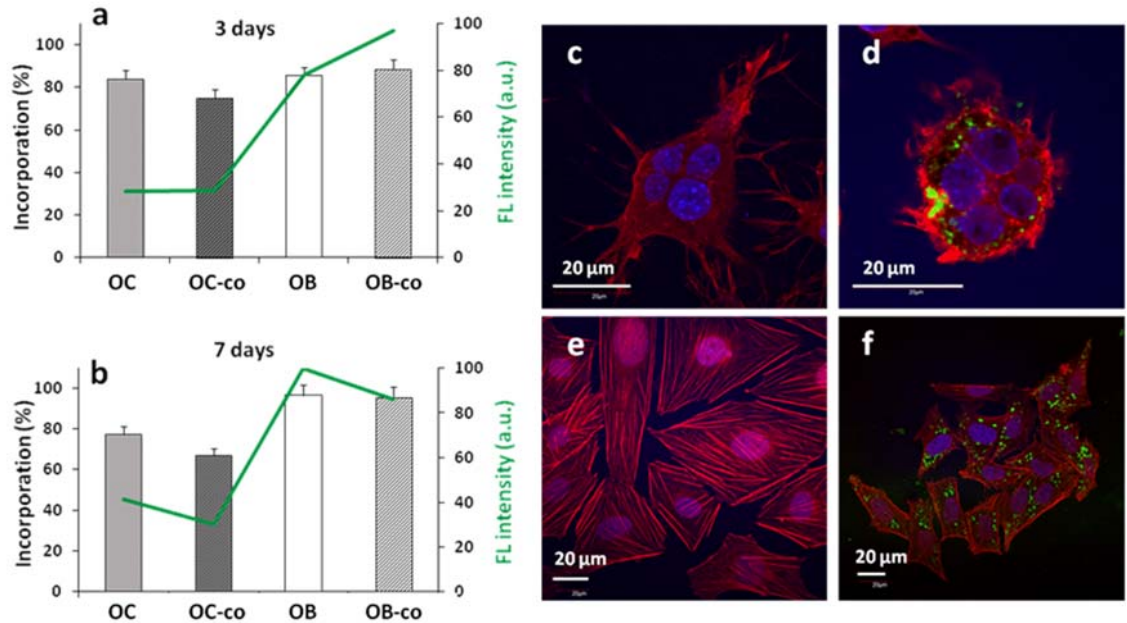

**Figure 6.** Incorporation of NanoMBG nanospheres labelled with FITC by osteoclasts and osteoblasts in monoculture (OC, OB) and in coculture with each other (OC-co, OB-co). Figures 6a and 6b show the flow cytometric analysis of intracellular incorporation of FITC-NanoMBG nanospheres (percentage of green fluorescent cells) and the fluorescence intensity (green line, arbitrary units) after 3 days (**a**) and 7 days (**b**). Confocal microscopy images of FITC-NanoMBG nanospheres incorporated by osteoclasts (**d**) and osteoblasts (**f**) in coculture with each other after 7 days. In the absence of nanospheres, control cocultures of osteoclasts (**c**) with osteoblasts (**e**) were carried out in parallel.

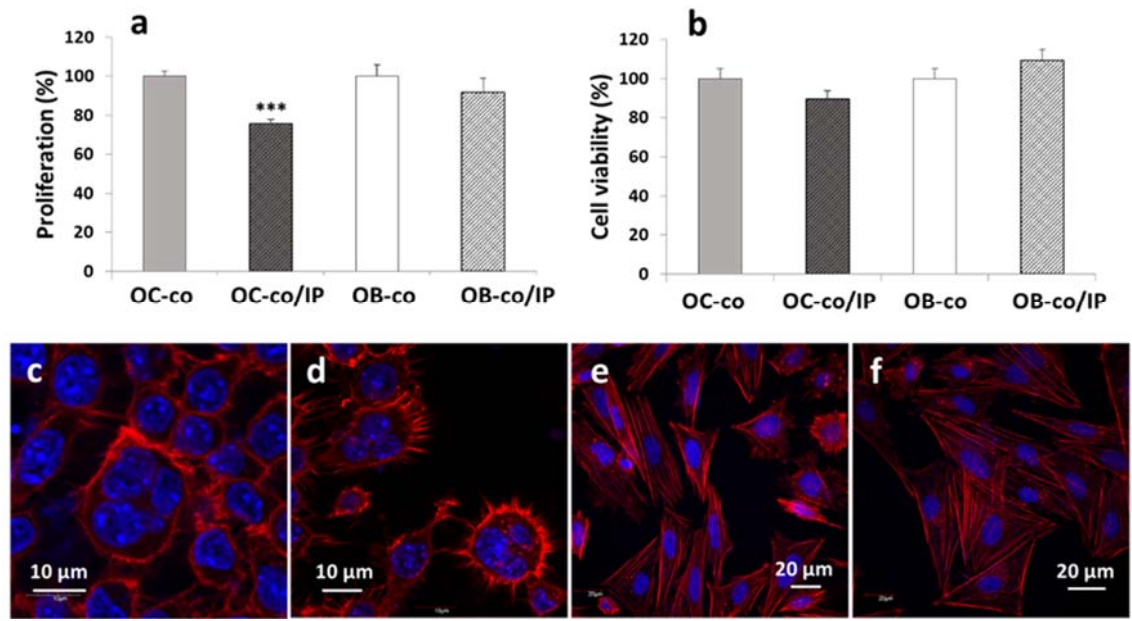

**Figure 7.** Cell number (**a**) and cell viability (**b**) of osteoclasts (OC) and osteoblasts (OB) in coculture with each other after 7 days with 50 μg/ml of either NanoMBG (OC-co and OB-co) or ipriflavone (IP)-loaded NanoMBG (Oc-co/IP and OB-co/IP) nanospheres. Cell morphology of OC-co (**c**), OC-co/IP (**d**), OB-co (**e**) and OB-co/IP (**f**).

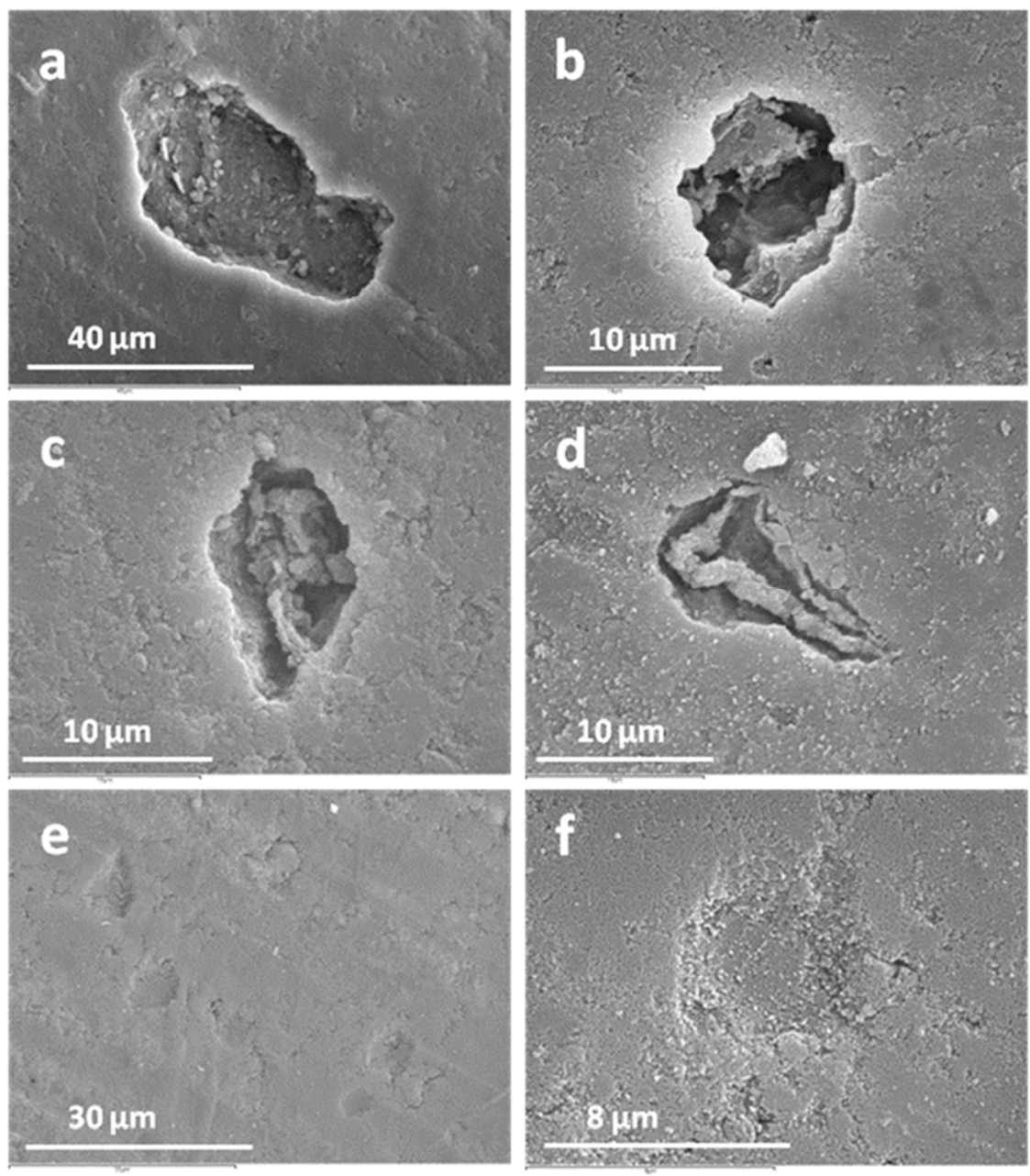

**Figure 8.** Scanning electron microscopy images of the resorption cavities left by osteoclasts after 7 days in monoculture (**a**), in coculture with osteoblasts (**b**), in coculture with osteoblasts in the presence of 50 μg/ml of NanoMBG nanospheres (**c** and **d**) and in coculture with osteoblasts in the presence of 50 μg/ml of NanoMBG-IP nanospheres (**e** and **f**).